\documentstyle[12pt,epsf]{article}

\newcommand{\bc}{\begin{center}}
\newcommand{\ec}{\end{center}}
\newcommand{\bd}{\begin{displaymath}}
\newcommand{\ed}{\end{displaymath}}
\newcommand{\be}{\begin{equation}}
\newcommand{\ee}{\end{equation}}
\newcommand{\ba}{\begin{array}}
\newcommand{\ea}{\end{array}}
\newcommand{\bea}{\begin{eqnarray}}
\newcommand{\eea}{\end{eqnarray}}
\newcommand{\bt}{\begin{tabular}}
\newcommand{\et}{\end{tabular}}

\newcommand{\bp}{\begin{picture}}
\newcommand{\ep}{\end{picture}}
\newcommand{\bfi}{\begin{figure}}
\newcommand{\efi}{\end{figure}}

\begin{document}


\title{\huge \bf {Initial Condition Model from Imaginary Part of Action
and the Information Loss Problem }}

\author{ 
H.B.~Nielsen ${}^{1}$ \footnote{\large\, hbech@nbi.dk} \\[5mm]
\itshape{${}^{1}$ The Niels Bohr Institute, Copenhagen, Denmark}}

\date{}

\maketitle

\begin{abstract}
We review slightly a work by Horowitz and Maldecena solving the 
information loss 
problem for black holes by having inside the blackhole - near to the 
singularity - a boundary condition, as e.g the no boundary proposal by 
Hartle and Hawking. Here we propose to make this boundary condition 
come out of our imaginary action model (together with Masao Ninomiya).
This model naturally begins effectively to set up boundaries - whether 
it be in future or past! - especially strongly whenever we reach to 
high energy physics 
regimes, such as near the black hole singularity, or in Higgs producing 
machines as LHC or SSC. In such cases one can say our model predicts miracles.
The point is that you may say that the information loss problem, unless you 
solve it in other ways, call for such a violation of time causality as in our 
imaginary action model!  
\end{abstract}

\newpage
\thispagestyle{empty}
\section{Introduction}
The information loss problem \cite{Samir} is essentially this:

1) From one point of view it seems that information falls into 
the black hole and 

2) the information comming out with the Hawking 
radiation seems not so easy to get correlated with the infallen 
information.

But then it is not easy to get as t'Hooft \cite{tHooft} would like: 
the blackhole 
is just 
a resonance: you scatter some particles and some other particles  come out 
connected 
by an ordinary S-matrix.

The point of the present talk is to look at the direction of solving 
this problem given by an article by Horowitz and Maldacena
\cite{Horowitz:2003he}, in which they propose to use  a fixing of the 
boundary conditions near or at the 
singularity inside the black hole. Let us immediately remark that 
such a direction of solving the problem is highly unconventional in the 
sense of having influence from the future, or should we say backward 
causation inside the black hole horizon. 

\section{Review of Mathurs putting of the trouble}
We heard Samir Mathur\cite{Samir} put the information loss problem 
roughly like this: 
In the usual Penrose diagram \cite{Penrose} for Black hole \cite{BH} the 
Horizon is a lightlike 
surface meaning in the Penrose diagram a lightlike directed line. 
We now imagine an extension of a curve of given external time t 
being  the Schwarzhild coordinate but not in the perfect Schwarzhild 
solution extended but in one a little bit more realistic model  including the 
very formation of the blackhole by some material falling in. Now we can 
find a surface, that is purely spacelike so that we could use it 
as description of a moment of time, a given special value of a time coordinate
being in the far outside simply the time $t$ but which would at the end 
be seperated by the horizon. 

The crux of the matter is that inside the horizon there are moment surfaces 
(space like surfaces, that could be taken as corresponding to a single value 
of a possible time coordinate) on which the information from the infallen 
material 
and the information correlated with the emitted Hawking radiation 
fall in widely different places. That is to say that the information about 
what falls in comes to one place at a certain ``moment of time'' while the 
Hawking radiation orginates at that moment of time from a far away region.
Thus it seems against the  principle of locallity to get the Hawking radiation
correlated with the incomming material. But this correlation is 
what is needed if it should be so that the black hole were simply functioning
as a normal resonance representing the scattering of the incomming material 
which then comes out again  as Hawking radiation.       

\section{Horowitz's and  Maldacena's ``solution''}
Since the problem of the information loss as here described is a 
matter of a problem with locality, one might think of having some 
form of violation of locality. Indeed  Horowitz and Madacena 
\cite{Horowitz:2003he} has a 
proposal
for solving the information loss problem by such a violation. In fact 
the idea of Horowitz and Maldacena is that there could exist a boundary 
condition 
imposed as a law of nature at or close to the singularity inside the 
black hole. Such a boundary condition at a time later than the time 
for which it is relevant means a restriction of the future, an arrangement 
that the future shall be in a special or restricted state, 
and thus it a priori opens up the 
possibilty for backward causation, meaning that the future influences the 
past. Now it were in the Horowitz and Maldacena considerations concerning the 
black hole only in the inside the horizon space time region that the backward
causation should take place. That should also be sufficient in order to 
solve the problem of the black hole information passing from the incomming 
matter to the outgoing Hawking radiation. In the outside the horizon region 
the reflection of the inside restriction at the singularity is that 
there comes a correlation between the infalling material and the outgoing 
Hawking radiation\cite{HR}. If we think of what happens on the above 
mentioned surface of events of a special moment of a certain time,
we can see that the future restriction can impose a correlation 
between the far away regions with respectively the infalling stuff and the 
Hawking radiation related degrees of freedom. Such a restriction at the 
singularity could - Horowitz and Maldacena also allude to as the possibility -
be due to the Hartle-Hawking no-boundary boundary condition\cite{noboundary}.  
To understand this idea of using a singularity based restriction 
``in the future'' to provide the needed correlation it may be needed 
to have in mind that the Hawking radiation emmited related degrees of freedom 
fall into the black hole and finally end up in the singularity. Thus these 
with Hawking radiation related degrees of freedom get via the future 
restriction related to the infalling stuff degrees of freedom and so 
finally the Hawking radiation comming out ends up related to the 
degrees of freedom of the infalling stuff.

\section{Miracles are called for, unless only say string stars (fuzz balls)}
In the foregoing section we saw that Horowitz and 
Maldecena\cite{Horowitz:2003he}  could help
on the information loss problem and thus make the black hole easier 
consistent with the developments as one usually expects it.  
I.e. we get as we expect the black hole to function as a normal 
resonance 
by introduction of  restrictions 
at the singularity. But if we first let in the possibility of restrictions on 
the future as some law of nature then we have opened up for the 
possibility of getting miracles into the theory. In fact if we 
have restrictions on what the future shall be then we should expect to see 
that some features of the state of the universe should be predestined to 
some special fate. Such arrangements would seem like miraclulously 
special arrangements. The restrictions in the future would only come about 
typically by happenings which a priori would look so strange that 
we would consider them miracles. Thus the type of theory proposed 
is a theory with miracles. Now it were in the case of solving the 
black hole problem of information loss only inside the black hole 
- inside the horizon - that were under the influence from future 
physics. Thus it were only inside the black hole that there were 
truly the need for the miracles, but if you allow them at all, it may
at the end be difficult to keep them away from the outside region. 

It should be mentioned that there may be other ways  - although it looks 
difficult - to solve the problem of information loss: In fact one can 
- and this is what stringtheory seems to deliver according to the talk 
of Samir Mathur\cite{Samir} - imagine that a genuine black hole never 
truly forms, but that the collapse stops - if not before - in the last
moment before a true black hole is formed. If truly a black hole were 
never formed, then of course the problems of information loss would not be 
relevant. In a way we can say that when as in Samir Mathurs picture
the size of the string or string theory material forming the potential 
black hole remains of the size of the Schwarchild radius \cite{BH} even when 
the
mass go so high that this Schwarchild radius has become very big, then 
it means that the black hole is not truly realized.

Such a keeping up the size of the stringtheory stuff to remain as 
big as the Schwarchild radius even when more and more stuff is being 
put on, can solve the information loss problem without need for any miracles.

\section{Ninomiya's and mine miracle model, imaginary part of action.}
The conclusion of the above discussion of the problem of black holes 
means that there is a call for a theory of the type with backward 
causation as suggested in the Horowitz and Maldacena 
article\cite{Horowitz:2003he} reviewed above.
This gives us the motivation and excuse for putting forward the 
model of Masao Ninomiya and myself \cite{own}. I made  an attempt to a popular 
presentation in a book of collections of talks on miracles \cite{ownpop} at 
AArhus University. Actually our ideas of influence from future are a bit 
related to old ideas of such influence mainly for the coupling constants
\cite{old} and for predestining humanity to make a new vacuum called 
the`` vacuumbomb''\cite{vacuumbomb}.

This model \cite{own}\cite{ownpop} is characterized by having in it a 
prediction 
of initial conditions in principle. It is even so that this in principle
predicted initial conditions are arranged so as to minimize a certain 
functional $S_I(history)$ depending on the history of the universe 
through all times from the beginning to the end, a functional being 
an integral over all space time, so that indeed it depends on both past 
and future. That is to say that the arrangement of the initial conditions
to appear in our model depends also on the future and thus will appear
as having prearrangement or backward causation in it. If for instance as we
suggest it in our model that  Higgs particles being produced will make 
$S_I(history)$ bigger than if they are not produced, then we should 
expect that there would be prearrangements occuring seemingly with 
{\em the purpose} of preventing the Higgs production. In fact one can 
approximately formulate the result of our predictions about the initial 
conditions by saying that they are adjusted so as to minimize the functional
$S_I(history)$. That is to say the history of the universe will 
in our model be approximately selected among all the histories possible
in accordance with the equations of motion as being that history which gives 
the smallest (i.e. most negative; it probably will be negative)
$S_I(history)$-value. 

This real quantity $S_I(history)$ which by being minimized 
determines the initial conditions is in our model actually the 
imaginary part of an -unusually - assumed {\em complex action}. 
That is to say our model consists actually in postulating that, 
contrary to what one usually takes it, the action for the development 
of the universe is fundamentally {\em complex}, i.e. 
of the form
\begin{equation}
S(history) = S_R(history) + i S_I(history).
\end{equation}         
This is to be understood that the parameters in the action 
- such as coupling constants and mass squares (in the case we consider 
of a quantum field theory, the standard model say) - are taken to 
be complex, while the fields (or the dynamical variables) are
taken as usual, i.e. real if they are real in the usual theory.

Since we now have an a bit unusual model with this complex action,
meaning complex couplings and masses, we shall in principle make the 
model precise by setting up - or rather choose - that formalism in which we 
want simply to insert the complex action instead of the usual real 
action. It is honestly speaking a further assumtion in our model to 
choose just into which expression  to insert the new complex action.
We choose to do it in a formalism using the Feynman-Dirac-Wentzel path 
way integral but in slightly special way: 

Usually one would use the Feynman-Wentzel-Dirac path way integration 
formalism  \cite{FH}\cite{Wentzel} 
to calculate a development ``matrix'' giving the time development from
one initial time $t_i$ to a final time $t_f$. Then the transition amplitude
from one initial state $|i>$ to a final state $|f>$ is given as a functional 
integral 
\begin{equation}
<f|U(t_i,t_f)|i> = \int \exp{\frac{i}{\hbar} S_{t_i --> t_f}(path)}
{\cal D}path,   
\end{equation}
where it is then understood that the action $S_{t_i--> t_f}(path)$ 
is the integral of the Lagrangian - taken of course for the $path$
being integrated over - over time from time $t_i$ to time $t_f$.
Also it is understood that the field values of the path at the end 
points in time $t_i$ and $t_f$, let us call them $\phi(t_i)$ and
$\phi(t_f)$ respectively, are to be integrated over with a weight 
given by the wavefuntion(al)s $<\phi(t_i)|i>$ and $<\phi(t_f)|f>$.

You would of course expect from the physical interpretation 
of the Feynaman path integral, that the weight of the contribution 
from the part of the integral where the fields at some time take 
the values in a certain interval should represent - in some way at least -
the probability for the fields having taken their values in that interval.
However, really the question as to what happens between the preparation 
of the state $|i>$ and the measurement of the final state $|f>$ cannot be 
answered because it would mean a new experiment to begin to measure on 
something in the intermediate time. It would be like in the Einstein Bohr 
discussion, if Einstein starts measuring through which of the  slits in the 
double slit experiment the particle goes. Nevertheless 
Aharonov et al. \cite{Aharonov} have discussed some weak measurements 
being performed in the intermediate time.

One could also for example make an expression for the average of 
an operator $O(t_f)$ at the time $t_f$ by means of the Feynman path way 
integral like this:
\begin{eqnarray}
<i|U(t_i --> t_f)^{\dagger} O U(t_i --> t_f)|i>& =&\\
= \int \exp{\frac{i}{\hbar}
S_{t_i --> t_f}(path)}O(\phi(t_f)) {\cal D}path \left ( \int \exp{
\frac{i}{\hbar} S_{t_i --> t_f}(path')} {\cal D}path' \right )^{\dagger}&,&   
\end{eqnarray}
where it is then to be explained that the boundaries for these two 
functional integrals at the end of time interval at $t_i$ should be integrated
over and weighted with the wave function $<\phi(t_i)|i>$ and its complex 
conjugate $<i|\phi(t_i)>$. It is also understood that the $path$ in the 
first factor and the $path'$ in the second integral are to be identified 
at the time $t_f$,
\begin{equation}
\phi(t_f)_{path} = \phi(t_f)_{path'}.
\end{equation}
 Finally the operator $O(\phi(t_f))$ should be understood as 
possibly depending even on the derivative of $\phi(t)$ 
derived w.r.t. to $t$ which is then identified with $t_f$.

Since in the usual case of the action being real the transition matrices 
as $U(t_f --> t_3)$ say is unitary we can even multiply in a product 
$U(t_f --> t_3)^{\dagger} U(t_f --> t_3) = 1$, and thus we also write 
\begin{eqnarray}
<i|U(t_i --> t_f)^{\dagger} O U(t_i --> t_f)|i>& =& \\
= \int \exp{\frac{i}{\hbar}
S_{t_i --> t_3}(path)}O(\phi(t_f)) {\cal D}path \left ( \int \exp{
\frac{i}{\hbar} S_{t_i --> t_3}(path')} {\cal D}path' \right )^{\dagger}&&,   
\end{eqnarray}  
where we now instead of at $t_f$ have the identification 
\begin{equation}
\phi(t_3)_{path} = \phi(t_3)_{path'}.
\end{equation}
That is to say that it does not matter for calculating the average 
of the operator $O(t_f)$ at $t_f$ whether we use the Feynman path integral
with a time interval going up to one $t_3$ or another, so that we could if we
would like take the choice of formulating it with taking $t_3$ to go say to 
infinity. 

If we wanted we could even replace the initial time $t_i$ state $<i|$
by inserting a projection operator $|í><i|$ and then putting also 
factor $1$ from unitarity on the initial time side of the formula.
In this way we could in the usual real action case obtain the expression 
for $<i|O(t_f)|i>$ of the form
  \begin{eqnarray}
<i|U(t_i --> t_f)^{\dagger} O U(t_i --> t_f)|i>& =&\\ 
\int \exp{(\frac{i}{\hbar}
S_{t_0 --> t_3}(path))}``(|i><i|)(\phi(t_i))''O(\phi(t_f)) 
{\cal D}path & *& 
\\ *\left ( \int \exp{
\frac{i}{\hbar} S_{t_0 --> t_3}(path')} {\cal D}path' \right )^{\dagger}&&,   
\end{eqnarray}
where $t_0$ is a time that can be anything provided it is earlier than
the time $t_i$ at which we have inserted the operator $``(|i><i|)(\phi(t_f))''$
which is to replace the initial state $|i>$ used at first. Since now 
this expression does not in the real action case depend on the 
times $t_0$ and $t_3$ provided they are outside the time interval 
$[t_i,t_f]$, we could choose them to anything we would like as long as 
these times $t_0$ and $t_3$ are  still outside. For instance we could 
take $t_0 = -\infty$ and 
$t_3 = + \infty$. We might even imagine as a slight generalization 
to insert several operators and think of replacing the special 
projection operator $``(|i><i|)(\phi(t_i)''$ by any operator $O_1(\phi(t_i))$
taken at the same time $t_i$ and thus write an expression like
\begin{equation}
 \int \exp{(\frac{i}{\hbar}
S_{t_0 --> t_3}(path))}O(\phi(t_i)) O(\phi(t_f)) 
{\cal D}path \left ( \int \exp{
(\frac{i}{\hbar} S_{t_0 --> t_3}(path'))} {\cal D}path' \right )^{\dagger}
,\label{inter}   
\end{equation}
as a suggestion for what we can use to extract information from a 
Feynaman path integral formulation. Note that 
this expression is quadratic in the Feynman path integral in the sence that 
it is a product of two Feynman path integrals, one with the dummy path 
being denoted $path$ and one complex conjugated with the dummy $path'$.

Now the idea is in the case of a complex  action $S=S_R + iS_I$ also 
to use this 
expression by postulating that this expression obtained by putting 
in combinations of operators into a Feynman path integral and then 
multiplying that by a complex conjugate path integral without the 
operator insertions to deliver expectation values for the to the operators 
associated quantities. That is to say we take it that our model 
is assumed to deliver the predictions obtained by being extracted 
from expressions of this type with $t_0$ and $t_3$ going to respectively
minus and plus infinity.

While in the case of usually assumed real action model the extension 
with the time intervals outside the interval $[t_i, t_f]$ used is
irrelevant, this is no longer true in the case of there being an imaginary
part of the action $S_I(path)$. So in our model it becomes important 
that we decide to use the whole time axis from the beginning to the end
of all times. 

It is strictly speaking an assumption being added into our model 
that we postulate just this type of interpretation of  our 
complex action model. We think, however, that just such an interpretation 
being based on using a Feynman-Wentzel-Dirac path integral extending 
a priori over all times from the beginning (big bang or whatever were the 
first moment, minus infinity likely) to the end of times is very reasonable.
After all, if we should somehow think of the path integral as the fundamental 
theory beyond quantum mechanics then it would not be so nice to choose the
time interval for evaluating the action to be put in the exponent 
of the path way integrand to depend on the choice of what we are to calculate,
or even worse on some arbitrary choice. In the case of the real action 
when the arbitrary choice of the times $t_0$ and $t_3$ does not matter
it would of course be o.k., but in the complex action where it would depend,
the natural assumption will be to take the maximal time interval 
over which to integrate to be the one to use.
  
We think that it is also very natural to associate the expectation value 
of an operator $O$ at a time $t$ to be associated with the 
path at the time t and those components of the fields the development 
of which are described by the path and associated to the operator $O$.
Thus we claim an interpretation of the form (\ref{inter}) to be 
quite reasonable.

\section{Significance of the imaginary part of the action}
Once we just have decided on an assumtion about the intepretation 
of our model using over all time Feynman path integrals - even 
without looking too much on the details alluded to in foregoing 
section of looking at expectaion values of operators and squaring 
the Feynman path integral - it should be rather unavoidable that 
only contributions to the Feynman path integral(s) from paths with 
the smallest (or rather most negative) imaginary action $S_I(path)$ will 
have much significance.
So it is almost obvious without much details that the history of the
universe that will effectively be the one realized must be characterized
by a very negative $S_I(history)$, the ``smallest'' $S_I(history)$.
That it will be so follows from the simple fact that the Feynman path integral
integrand has the factor $\exp{-S_(path)}$, which will only be dominant when
the imaginary part of the action $S_I(path)$ is very negative. 

Really we should have in mind that most likely the very long time intervals 
after $t_f$ ``the future'' and before $t_i$ ``the past'' will give big 
contributions to the imaginary part of the action, that will suppress most 
possible developments so much that essentially almost only one development,
one history of the universe, comes to dominate. The best may actually be to 
think of performing a classical approximation. It is wellknown how in the 
Feynman path integral formulation one obtains the classical approximation as 
a saddle point approximation to the functional integral. Pedagogically 
- and to avoid a complicated discussion to extract at all a 
classical approximation, which is not so obvious at first - we may 
assume the imaginary part (at first) to be small. Then we would 
in first approximation be allowed to think upon  the Feynman path integral as 
being in the classical approximation given by the saddle point 
contributions calculated at first as if we only had the real part
$S_R(path)$.This would mean quite usual classical solutions to the 
equations of motion would be all that would contributes in this approximation.
But even a small imaginary part compared to the real part could give 
enormously big factors of the form $\exp{-S_I(history)/\hbar}$. We shall in 
fact 
not forget that in practice one expects $\hbar$ to be very small, so that for 
this reason already the exponent gets huge. When we think about that we have 
to do with integrals over time regions of the size of the whole lifetime
of the universe, these imaginary action values for the whole life span 
of the unviverse will easily suppress almost all but one single classical 
solution. That is to say that even a in some sense small imaginary part 
would be far sufficient to drastically select almost only one sadle point 
 contribution to survive being of significance.

We thus arrive at the first approximation description of the prediction
of our model namely in a classical approximation: 

First imagine calculate
all the classical solutions using just the real part of the action. This 
delivers a set of all the classical solutions. Then calculate for each of 
these soltions (in practice of course we cannot do that, but think of 
it in principle), these possible histories $history$, the imaginary part 
of the action $S_I(history)$. Then our model predicts that just that history 
$reahist$ which gets  recognized as the realized one, the one 
that truly happens, is the one for which $S_I(reahist)$ is {\em minimal}.

This is what we could refer to as the formula ``for the will of God''
being
\begin{equation}
S_I(reahist) \hbox{ shall be {\bf minimal}}.
\label{Godwill}\end{equation}  

It should be had in mind that this type of determination of the 
initial state to be realized depending on an integral $_I(history)$
which invloves all times, means that the way the universe developments has 
been 
started in a way depending on what could happen or not happen at much later 
times. But that then means that it would be like miracles, namely as if 
things have been prearranged in a statistically unexpected way, so as to just 
arrange that especially negative contributions to $S_I(hisory)$ get 
favoured by the selection of what really happens.  I.e.
negative imaginary part of the Lagrangian $L_I(history(t))$ (where 
$history(t)$ means the state and development derivatives in the history 
$history$ at time equal to $t$) gets favoured to be big. One of our 
speculations to be discussed 
in  section \ref{Higgs} is that production of Higgs particles should 
cause a relatively huge positive  contribution to $L_I$ so that histories 
leading to 
Higgs production become disfavoured.

\subsection{Really strong assumption, if we take action real}
First let us, however, now give an argument that, it would be very nice 
estetically to have the action  not being restricted to be real, 
but rather to also have an imaginary part. Indeed we may simply think of the 
Feynmann-Wentzel path integral - even written only very abstractly without
going in detail - written in the form
\begin{equation}
\int exp(\frac{i}{\hbar} S(path)) {\cal D}path.
\end{equation} 
Now we come with the remark that it is completely obvious that the integrand 
$exp(\frac{i}{\hbar} S(path])$ is {\em complex}. There is namely 
even simply an $i$ present as an over all factor in the exponent. 
If we therefore take the point of view that the most fundamental and 
important quantity is the integrand rather than say the action itself,
then we could say: if something should be assumed to be real rather 
than complex, then it should be this most fundamental quantity 
that should be assumed to be real.  

To take the integrand  to be real would of course be completely unacceptable, 
if one would have any connection to the usual theory. So the natural 
possibility is that there is {\em no restricton to reality at all},
so that both $S(path)$ and the integrand are allowed to be complex.

This argumentation may also be made more concrete by imagining that 
one would find some theory behind the Feyman Wentzel path way 
descripton of quantum mechanics, i.e. some model from which one derives 
quantum mechanics and arrive to a Feynman Wentzel path way formulation.
Then if one would hope for the usual theory with the {\em real} 
action it would be a very delicate mechanism that would be needed to 
ever get the 
integrand
become just a quantity of norm unity - as is what the real action means-.
For example we attempted\cite{bled} such a  derivation of quantum mechanics 
in the 
path way formulation and indeed did not at first find any reason 
why the integrand should be of norm unity.

\subsection{How to hide roughly the imaginary part of action}
At first it would look that our model with 
the complex action would lead to too many prearranged happenings to
agree with what we observe; there would be too many ``miracles''
or ``antimiracles''(repectively good or bad a priori unlikely 
events). Now, however,  we have found some mechanism that might indeed 
help to reduce the predicted number of such at first unlikely arrangements
in practice. Let us here in this discussion already accept the above mentioned 
classical approximation that we just have the effect of the imaginary part of
the action, $S_I(path)$ in our model so that it just delivers the formula
(\ref{Godwill}) to select the realized solution to the classical equations
of motion. 

The important point that brings down dramaticly the number of strange 
events, miracles or anti miracles, is that with the restriction from the 
equation of motions it is made troublesome to make too many miracles.
If the initial state so to speak has to be adjusted to make certain special 
event at one moment of time then the degrees of freedom of this initial 
state are partly fixed by this arrangement and there is less freedom to
adjust them to make - arrange for - more miracles. Thus it looks that the 
longer time the universe exists the more competition there will be 
about getting arrangements to each indivdual era of times. Only 
the ``miracles'' or ``anti miracles'' in the human history has a good chance 
to be spotted by poeple, and even then probably mainly the ones close to 
our own times, if we shall get aware of them. But since the universe has 
an age of the order of 13 milliard years already alone, the fact 
that there were $10^{8}$ human age periodes in even just the certainly 
existing time periode of for the universe, each arrangement would have to 
be shared by at least these $10^8$ periodes. Actually we believe 
from consideration of an action ansatz analogous the real part $S_R$
already known from phenomenology of the equations of motions, that it is 
likely that high energy scale physics contributes the most. When we think
of the action as being written as a four dimensional space time integral 
$\int {\cal L} d^4x$ with the Lagrangian density having dimension mass to the 
fourth power, it should be obvious that in order to get a big contribution 
to the action from space time volume of a given size, we should involve 
physics 
with so high energies (per particles) involved as possible. Now the universe 
were of 
smaller size in the time shortly after big bang and the time scales of the 
eras were smaller so that this presumed higher contribution from the high 
 energy scale being high is partly compensated for by smaller space time 
volume.

Nevertheless it is highly possible that a major contribution could 
have come to the imaginary part $S_I(history)$ from the era of inflation.
One could even imagine that the as slow roll problem presented 
 phenomenological call for a somewhat suspicially long time during 
which the inflaton field remained in a special region could be one of the
``miracles'' in our model. That should mean that because a special value 
for the inflaton field would give especially numerically high but negative 
imaginary part $L_I$ this value of the inflaton field would be (pre)arranged
to be taken on over the biggest possible space time volume, 
the specially favoured 
value giving the very negative $L_I$. That might indeed favour what would
look like a miraculously long stand in the inflation state.

If indeed some contribution from the time around big bang  might dominate 
numerically $S_(history)$ then the initial state would be dominantly 
arranged to make the dominant imaginary action contribution possibly 
most negative and then there would be less freedom of adjusting to 
make miracles at other times in the development of the universe. The point 
of course is that if first the initial state of the universe has been 
adjusted to give the smallest or most negative $S_(history)$
in some era near big bang like the inflation era, then there is less 
freedom to arrange miracles in later time. The equations of motion will
namely determine what happens later once it has been determined
with some``purpose'' related to the inflation era, say.

So the hypotesis of a dominant era different from our own concerning the 
imaginary action $S_I(history)$ would help to reduce the number and degree 
of remarkableness of the miracles or anti miracles to be predicted to 
occur to day.

\section{Our prediction of the failure of LHC or other Higgs producing 
machines}
\label{Higgs}

It is the suggested to be the major experimental test of our model of 
there being an 
imaginary part for  the action, that production of many Higgs particles
should be suppressed in the sense that machines destined to make big 
amounts of Higgs particles should have bad luck.  We shall therefore now 
review 
this part of our model, i.e. the arguments for this bad luck comming out 
of our model for the big Higgs producing machines:

The most natural assumption about the order of magnitude of the 
imaginary part would be that we take the various coefficients in the 
expression for the Lagrangian density such as coupling constants and 
mass squares ($m_h^2$ say) should have rather random phases of order 
unity. That would 
imply that we would expect the imaginary part $S_I$ and the real part $S_R$
of the action to be of similar order of magnitude for some random 
field development. For the special development, which the universe 
should perform in our model, and which is selected by minimizing the imaginary 
part it could be different. There is, however, one coefficient in the 
Standard Model Lagrangian density which requires a special consideration 
concerning the 
order of magnitude of real versus imaginary part, namely the mass square 
of the Higgs particle. The special point about this mass square of the Higgs 
particle is that it is a very wellknown {\em mystery} why the mass square 
of the Higgs particle defined in a renormalized way is so enormously 
small compared to the magnitude,which we would expect to have, namely the 
fundamental 
mass square scale of physics supposedly the Plack mass squared. This mystery 
we may call the `scale problem'' - why so different scales ?! -. It would 
be even more a problem, if one would like to have a unified gauge theory 
like $SU(5)$ or $SO(10)$ or the like, since the unifying scales 
would also be far away from the Higgs mass scale - one would then even have the
doublet triplet seperation problem-. It is really this scale problem
that manifests itself by giving rise to the hierarchy problem: How to 
avoid that by each new perturbative correction to the 
Higgs VEV or Higgs renormalized mass quadratic divergences -
supposedly cut off  by some fundamental physics at the Plack scale -
do not reshuffle the Higgs mass (square) by enourmous amounts
recalling the scale problem mystery order by order again.

The part of the Higgs mass square coefficient for which we know 
the order of magnitude phenomenologically is the {\em real} part
$m_H^2|_R$ of this coefficient to the Higgs field squared $m_H^2$
in the Lagrangian density ${\cal L} = ...+ m_H^2/2 * \phi_H^2(x)+ ...$,
where we have split up the coefficient
\begin{equation}
m_H^2 = m_H^2|_R + i* m_H^2|_I.
\end{equation}
    
But now, if it is so mysterious, why the real part $m_H^2|_R$ is so 
small compared to the Plack scale mass square (the square of the 
Planck mass), and we do not really understand yet the true 
mechanism for it being so small, then how can we know whether this
``mysterious'' mechanism works to also make the imaginary part 
of the Higgs mass square $m_H^2|_I$ surprisingly small? Very likely
it will actually {\em not} make also the imaginary part small, because what is 
truly what is small concerning the real part is not simply the bare 
real part, but rather the by several corrections modified - i.e. relative 
to that 
dressed or renormalized -  real part. But renormalizing the imaginary 
part would likely be a quite different story so that a mysterious finetuning
tuning the renormalized real part to be exceptionally small compared to the 
a priori expectation, the Planck mass square, would most likely not
hit to make the imaginary part small. So we actually expect the 
imaginary part $m_H^2|_I$ still to be of the order of the Planck mass square.
But now from the point of view of  the typical energy scales of say the 
LHC accelerator 
- a few TeV - the mysterious small Higgs mass and 
thus the real part $m_H^2|_R$ is of  a rather normal order of magnitude,
while an imaginary part of Plack scale size would seem enourmously big!
This then means that as soon as the imaginary part of the square of the 
Higgs mass comes in, it will completely dominate the present day 
contributions to the imaginary action. Now in the experiments 
we have so far studied, not even seeing the Higgs yet at all, the couplings 
and masses relevant have only been what came out of the dimensionless 
couplings/coefficients  in the Standard model Lagrangian density 
and the {\em Higgs vacuum expectation value}. The latter is determined from
 the real part $m_H^2|_R$ and thus the imaginary part would get so far 
no influence. There would  as long as no Higgses 
are truly produced only be a constant vacuum contribution from the term 
$ m_H^2|_I |\phi(x)|^2 $ to the imaginary Lagrangian density ${\cal L}(x)$.

Only when the Higgs field is modified relative to its uusual vacuum 
value VEV= $<\phi_H>$ will the imaginary part come into play in a variable 
way. But that is 
typically the Higgs production and the existence of genuine Higgs particles 
flowing arround. We therefore expect that it is the flowing arround 
of produced Higgses, that will contribute the very likely very huge 
contribtuion to the ${\cal L}(x)$ and thus to $S_I(hitory]$. Now presumably
the appearance of geuine Higgs particles flowing arround is presumably 
a positive contribution to the imaginary action so that it would be 
disfavoured in the selection of the truly realized solution to the equations 
of motion to have Higgses arround; if it were namely instead very favoured 
we should already have Higgses all over. 

Thus we now expect that it would make the imaginary part of the 
action $S_I(history)$ appreciably bigger (less negative) if in the 
history of the universe many Higgses come to exist, thus accelerators 
like SSC, the Tevatron, and the LHC producing Higgses should 
preferably for minimizing $S_I$ be avoided by not comming to work 
or quckly be stopped again once working. One should of course also then 
expect that cosmic rays should miraculously or somehow from the 
initial conditions of the universe have been arranged to produce 
as few Higgses as can easily be organized without spoiling too 
much the possibilities for the appropriate miracles in other eras 
so that their negative $S_I$-contributions are not too much reduced from what 
they can maximally be. But we humans have little understanding of 
how much cosmic rays there would have been under slightly different 
choices of the initial conditions, so we do not know if there should be 
a fine tuning of the initial conditions so as to make few or many cosmic
radiation particles. Contrarily we have, however, good understandings of, 
that when one has built about the quarter of the tunnel of the 
planned SSC (= superconducting super collider)\cite{SSC} in Texas, then there 
is apriori a very high expectation that there should soon be produced 
a lot of Higgs particles (if it exist at all of course as we assume here).
Then it were really like an anti-miracle, when the Congress stopped the 
machine 
from being built and let the tunnel be only used for champignon growing 
or the like.

In our model we actually take this somewhat surprising fate of bad luck 
for the great SSC-project as an anti-miracle confirming our model.
Also the accident of a bad connection stopping for soon 
a year the LHC just, when it were about to start functioning, we take as
a symptom of our model! It should be stressed that our predictions 
about bad luck for Higgs producing accelerators were made after the bad fate of
SSC, but {\em before} the accident at LHC that have delayed it 
by now soon a year!

.  
\section{Conclusion}

The main point of the presnt article were to call attention to that 
by the ideas of Horowitz and Maldacena \cite{Horowitz:2003he} for solving 
the problem with black holes 
of correlating the infall information with the outgoing Hawking radiation
a {\em backward causation} theory is called for. In competition with 
for instance 
the Hartle Hawking no-boundary postulate\cite{noboundary} replacing 
the singularity with 
a special condition - say no boundary - thereby imposing ``final conditions''
leading to backward causation we presnted ``the imaginary part of action 
model'' by Masa Ninomiya and myself\cite{own}. The crux of this matter 
were, 
really as explained in the talk by Samir Mathur, the problem of getting the 
information from the infalling stuff into the black hole transfered to the
outgoing Hawking radiation. This  is a problem of causality - like the problem 
of 
tranfering information from one place to another place faster than with 
speed of light-. The problem  would therefore possibly be avoided, if we
have a theory with backward causation, so that future can influence past and
therefore no causality principle can be truly valid. For phenomenological 
reasons 
it is of course needed that under ``normal'' conditions the amount of 
backward causation - or as we also refered to cases of backward causation,
miracles or anti miracles - should be seldom. This is indeed the case 
both by thinking of Hartle Hawking no-boundary (mainly showing up in 
black holes, which are phenomenologically badly known) and in our 
``imaginary part of action model'', in which it is though needed a 
somewhat speculative argumentation to argue that the cases 
of backward causation get so seldom as needed for agreement with dayly
life experience. We think, however, that there {\em is} a good chanse
that the restriction from the history of the universe having to obey the 
(classical) equations of motion (at least approximately) could impose 
so strong restrictions on the amount of backward causation or miracles 
or anti miracles that it would not disagree with present knowledge.
In this way we want to claim that our model is viable so far.


\end{document}